\documentclass[12pt]{article}
\usepackage{eurosym}
\usepackage{graphicx}
\usepackage{subcaption}
\usepackage[T1]{fontenc}
\usepackage{indentfirst}
\usepackage[margin=0.6in,nomarginpar]{geometry}
\usepackage[final]{hyperref}
\usepackage{amsmath}
\usepackage{hyperref}
\usepackage{cite}
\usepackage{xcolor}
\usepackage{subcaption}
\usepackage{caption}
\usepackage{amssymb}
\usepackage{multirow}
\usepackage[table]{xcolor}
\usepackage{orcidlink}
\usepackage{float}
\usepackage{bookmark}
\usepackage{hyperref}
\usepackage{graphicx}
\usepackage{physics}

\definecolor{revcolor}{RGB}{180,0,0}

\hypersetup{
colorlinks=true,
linkcolor=blue,
citecolor=blue,
filecolor=magenta,
urlcolor=blue
}

\begin{document}

\title{Dunkl--Pauli Oscillator in an Aharonov--Bohm Flux:
Restriction of Admissible Angular Sectors and Thermodynamic Signatures}

\author{Ahmed Tedjani\,\orcidlink{0000-0002-9590-4082}$^{1}$\thanks{%
Email: \texttt{ahmed.tedjani@univ-constantine3.dz}}
\and
Boubakeur Khantoul\,\orcidlink{0000-0001-9012-4864}$^{1,2}$\thanks{%
Email: \texttt{boubakeur.khantoul@univ-constantine3.dz}}
\\[6pt]
$^{1}$Department of Process Engineering, University of Constantine~3
--Salah Boubnider\\
25016 Constantine, Algeria\\
$^{2}$Theoretical Physics Laboratory, Department of Physics,
University of Jijel, Algeria}

\maketitle

\begin{abstract}
We study a two-dimensional Dunkl--Pauli oscillator in the presence of an Aharonov--Bohm (AB) flux.  The combination of reflection symmetry (via Dunkl operators) and a topological gauge field imposes a nontrivial constraint on the admissible quantum states: the regularity condition on radial wave functions, together with the matching conditions at the flux tube, leads to a compatibility relation $\nu_1+\varepsilon\nu_2=0$ and forces the emergence of a lowest angular quantum number~$\ell_0$.  As a result, the admissible angular sectors are restricted rather than the spectrum being merely shifted in energy.

Using the exact spectrum, we construct the canonical partition function and derive closed-form expressions for the internal energy, entropy, and heat capacity.  The thermodynamic quantities directly reflect this spectral constraint: the low-temperature behaviour is governed by~$\ell_0$, and the heat capacity exhibits a flux-controlled Schottky-like peak.  The Dunkl parameter~$\nu$ plays a dual role: within a fixed sector it shifts the energy scale, while globally it controls the flux threshold that determines which sectors are admissible.  At high temperatures the classical two-dimensional oscillator limit is recovered.

Our results demonstrate that the interplay between Dunkl symmetry and AB flux qualitatively modifies the set of admissible states, with observable thermodynamic signatures.
\end{abstract}

\section{Introduction}
\label{sec:intro}

The study of low-dimensional quantum systems has revealed a wide
range of phenomena with no classical analogue, particularly when
topology and symmetry play a central role
\cite{Leinaas1977,Wilczek1982}.  In two spatial dimensions, the
interaction between charged particles and gauge fields depends not
only on local forces but also on global properties of the
configuration space.  A paradigmatic example is the Aharonov--Bohm
(AB) effect, in which a magnetic flux confined to an inaccessible
region modifies the quantum phase of a particle moving in a
field-free domain \cite{Aharonov1959,Aharonov1961,Olariu1985,%
Peshkin1989}.  This phenomenon demonstrates that the vector potential
has direct physical significance, influencing measurable quantities
such as energy spectra and interference patterns, as confirmed
experimentally by Chambers \cite{Chambers1960} and Tonomura et al.\
\cite{Tonomura1986}.  The AB effect is intimately related to
geometric phases and gauge invariance \cite{Berry1984} and continues
to play an important role in modern quantum theory.

For particles with spin, the appropriate framework is provided by
the Pauli Hamiltonian \cite{Pauli1927}, which describes the coupling
between the intrinsic magnetic moment and an external electromagnetic
field.  As the non-relativistic limit of the Dirac equation
\cite{Dirac1928a}, the Pauli equation captures spin precession and
the Zeeman interaction.  In two-dimensional geometries, the combined
action of spin interactions and magnetic flux leads to nontrivial
spectral structures, particularly when boundary conditions and
rotational symmetry are taken into account \cite{Hagen1990}.  Such
systems are relevant in mesoscopic physics and quantum nanostructures,
including quantum rings and two-dimensional electron gases, where
topological phases can be experimentally observed
\cite{vonKlitzing1980,Tsui1982,Beenakker2008}.  The Pauli equation
in the presence of an AB flux has therefore been studied in various
contexts \cite{Bouguerra2008,Bouguerne2024}.

Beyond conventional quantum mechanics, deformations based on Dunkl
operators have attracted considerable interest in mathematical
physics.  Introduced by Dunkl \cite{Dunkl1989,Dunkl1991} and further
developed in subsequent works \cite{Rosler2003,Dunkl2008,Dunkl2014},
these operators incorporate reflection symmetries directly into
differential operators.  
In the context of integrable many-body systems, Dunkl operators provide a powerful framework for solving the Calogero--Moser model \cite{Rosler2003,Chalykh2025}: they map the singular $1/r^2$ interactions to a set of decoupled harmonic oscillators, thereby encoding particle statistics directly into the kinetic structure \cite{Chalykh2019,FeiginHakobyan2022}. Recent work has further extended this framework to generalized Calogero models with reflection group symmetry \cite{Hakobyan2025}.
The kinetic term of the resulting Hamiltonian contains discrete
reflection operators characterised by deformation parameters.  The
Dunkl formalism has been applied to a variety of quantum systems,
including Dirac and Klein--Gordon equations with Dunkl derivatives
\cite{Mota2021a,Mota2021b,Hamil2024DunklKG}, Dunkl--Pauli systems
in magnetic fields \cite{Bouguerne2024,Benchikha2025,Salazar-Ramirez2026},
and noncommutative phase-space extensions \cite{Hassanabadi2023}.
Recent studies have extended the Dunkl framework to the Schr\"odinger
equation in higher dimensions \cite{Hamil2025DunklSchr}, the Morse
potential \cite{Hamil2025Morse}, time-dependent Dunkl systems
\textcolor{blue}{\cite{Lutfuoglu2025TimeDep}},
and via path integral methods \textcolor{blue}{\cite{Hamil2024PathIntegral}}.

In parallel, the thermodynamic properties of Dunkl-deformed quantum systems have been investigated for various models, including \cite{DOI:10.1140/epjp/s13360-022-03055-1, DOI:10.1140/epjp/s13360-022-03463-3, Bouguerne2024, DOI:10.1088/1742-5468/acd106}, providing essential context for the present statistical mechanical analysis.

Despite these advances, the combined effect of an AB flux and Dunkl
reflection symmetry has remained largely unexplored.  Existing works
treat either topological phases in conventional Pauli systems or Dunkl
deformations in the absence of singular gauge fields.  The interplay
between a point-like magnetic flux, which breaks rotational symmetry
in a topological way, and the discrete reflection operators, which
introduce parity-dependent effective potentials, raises a fundamental
question: \emph{does the singular AB flux impose additional
constraints on the allowed reflection sectors and angular states?}

In this work we answer that question affirmatively.  We investigate
the stationary Dunkl--Pauli equation in two dimensions in the
presence of an AB flux.  The Hamiltonian is obtained by replacing
the canonical momentum operators with Dunkl operators in the Pauli
Hamiltonian---referred to hereafter as the \emph{Dunkl--Pauli
Hamiltonian}.  After separating the angular and radial parts, we
solve the eigenvalue problem with proper self-adjoint boundary
conditions at the flux tube.  The key result is that the regularity
condition $K>-1$, required for square-integrable radial wave
functions with respect to the Dunkl measure, does \textbf{not}
merely shift the energy levels; it restricts the set of admissible
angular quantum numbers.  Specifically, the combined effect of the
AB flux~$\vartheta$ and the Dunkl parameters $(\nu_1,\nu_2)$ forces
a lowest admissible angular quantum number~$\ell_0$ that depends on
both the flux and the reflection sector.  This $\ell_0$ emerges
from the compatibility condition $\nu_1+\varepsilon\nu_2=0$ (with
$\varepsilon=\pm1$ labelling the reflection parity) and directly
modifies the ground-state energy~$E_0$.  Consequently, not every
angular momentum state that would be allowed in the pure Dunkl
oscillator or in the pure AB oscillator survives when both are
present simultaneously.

To probe the physical consequences of this restriction, we turn to
thermodynamics.  The canonical partition function is constructed by
summing only over admissible states, with $\ell_0$ entering through
the ground-state energy.  We compute the internal energy, entropy,
and heat capacity and show that the flux and deformation leave clear
imprints---most notably in the low-temperature behaviour, where
ground-state dominance amplifies the effect of~$\ell_0$, and in the
Schottky-like peak of the heat capacity, which is controlled
by~$\vartheta$.  The Dunkl parameter~$\nu$ plays a dual role: within
a fixed sector it shifts the energy scale, while globally it controls
the flux threshold that determines which sectors are admissible.  At
high temperatures, the system approaches the classical oscillator
limit, as expected.  The thermodynamic analysis thus serves as a
concrete demonstration of how the spectral constraint manifests in
observable quantities, with potential signatures in mesoscopic ring
experiments.

This study complements our previous analysis of the time-dependent
case \cite{Khantoul2026} by focusing instead on stationary states
and their thermodynamic implications.  The novelty lies in showing
that the AB flux does not merely shift the spectrum but restricts the
admissible angular sectors through the emergence of~$\ell_0$, a
direct consequence of the interplay between Dunkl reflection symmetry
and the topological gauge field.

The paper is organised as follows.  Section~\ref{sec:hamiltonian}
presents the Dunkl--Pauli Hamiltonian with the AB flux, separates it
into radial and angular parts, establishes the Hilbert-space
structure and self-adjointness conditions, and derives the matching
conditions and compatibility relation leading to~$\ell_0$ in full
detail.  Section~\ref{sec:thermo} constructs the partition function
and analyses the thermodynamic properties, including a detailed
physical interpretation of the $\ell_0$ constraint.
Section~\ref{sec:realization} discusses possible physical
realisations of the Dunkl deformation.
Section~\ref{sec:conclusion} concludes with a summary and outlook.

\section{Dunkl--Pauli Hamiltonian with Aharonov--Bohm Flux}
\label{sec:hamiltonian}

We consider a spin-$\frac{1}{2}$ particle confined to the $(x,y)$
plane and subjected to a static harmonic oscillator potential in the
presence of a magnetic field.  Its stationary states are governed by
the time-independent Pauli equation
\begin{equation}
\left[
  \frac{1}{2M}\boldsymbol{\pi}^{2}
  -\frac{e}{2M}\sigma_{z}B_{z}(\mathbf{r})
  +\frac{1}{2}M\omega^{2}(x^{2}+y^{2})
\right]\psi(x,y)=E\,\psi(x,y),
\label{eq:pauli}
\end{equation}
where $\boldsymbol{\sigma}=(\sigma_{x},\sigma_{y},\sigma_{z})$ are
the Pauli matrices and $\boldsymbol{\pi}=\boldsymbol{p}-e\mathbf{A}$
is the gauge-invariant momentum,
\begin{equation}
\boldsymbol{\pi}=\bigl(-i\partial_{x}-eA_{x},\,-i\partial_{y}-eA_{y}\bigr),
\end{equation}
with $e$ the particle charge and natural units $\hbar=c=1$.  The
two-component spinor is $\psi=(\psi_{1},\psi_{2})^{T}$.  Although
the identity $(\boldsymbol{\sigma}\cdot\boldsymbol{\pi})^{2}
=\pi^{2}-e\sigma_{z}B_{z}$ reduces the kinetic term to a scalar
form, the spin-dependent Zeeman contribution remains explicitly
present and plays a crucial role in the matching conditions at the
singular flux; the system therefore cannot be reduced to a purely
scalar Hamiltonian.

\subsection{Aharonov--Bohm Configuration}

To incorporate the AB effect \cite{Aharonov1959,Aharonov1961}, we
consider a magnetic flux~$\vartheta$ confined to an infinitely thin
solenoid along the $z$-axis.  The magnetic field is localised at the
origin:
\begin{equation}
B_{z}(\mathbf{r})=\vartheta\,\delta^{(2)}(\mathbf{r}),
\label{eq:bfield}
\end{equation}
which in polar coordinates reads $B_{z}=\tfrac{\vartheta}{r}\delta(r)$.
In the Coulomb gauge the associated vector potential takes the
azimuthal form
\begin{equation}
e\mathbf{A}=-\frac{\vartheta}{r}\mathbf{u}_{\varphi}
=\frac{\vartheta y}{x^{2}+y^{2}}\,\mathbf{i}
-\frac{\vartheta x}{x^{2}+y^{2}}\,\mathbf{j}.
\label{eq:vectorpotential}
\end{equation}

\subsection{Dunkl Deformation of the Momentum Operator}

The Dunkl formalism introduces reflection operators into the kinetic
structure \cite{Dunkl1989,Dunkl1991}.  We replace the canonical
momentum by the Dunkl momentum operator
\begin{equation}
p_{j}=-iD_{j},\qquad
D_{j}=\frac{\partial}{\partial x_{j}}+\frac{\nu_{j}}{x_{j}}(1-R_{j}),
\label{eq:dunklderiv}
\end{equation}
where $\nu_{j}>-\tfrac{1}{2}$ are real deformation parameters and
$R_{j}$ are reflection operators satisfying $R_{j}f(x)
=f(\dots,-x_{j},\dots)$ and $R_{j}^{2}=1$.  The Dunkl Laplacian is
\begin{equation}
\triangle_{D}=\sum_{j=1}^{2}D_{j}^{2}
=\sum_{j=1}^{2}\left(
  \frac{\partial^{2}}{\partial x_{j}^{2}}
  +\frac{2\nu_{j}}{x_{j}}\frac{\partial}{\partial x_{j}}
  -\frac{\nu_{j}}{x_{j}^{2}}(1-R_{j})
\right),
\label{eq:dunkllaplacian}
\end{equation}
and the operators satisfy the deformed commutation relations
\begin{equation}
[x_{i},D_{j}]=\delta_{ij}(1+2\nu_{j}R_{j}),\qquad
[D_{i},D_{j}]=[x_{i},x_{j}]=0.
\label{eq:algebra}
\end{equation}

\subsection{Explicit Form of the Dunkl--Pauli Hamiltonian}

Substituting Eqs.~\eqref{eq:dunklderiv} and
\eqref{eq:vectorpotential} into Eq.~\eqref{eq:pauli} yields
\begin{equation}
\begin{aligned}
H={}&-\frac{1}{2M}\triangle_{D}
     +\frac{1}{2M}\frac{\vartheta^{2}}{x^{2}+y^{2}}\\
&+\frac{1}{2Mi}\!\left[
  \frac{2\vartheta x}{x^{2}+y^{2}}\frac{\partial}{\partial y}
 -\frac{2\vartheta y}{x^{2}+y^{2}}\frac{\partial}{\partial x}
 +\frac{2\vartheta x}{x^{2}+y^{2}}\frac{\nu_{2}}{y}(1-R_{2})
 -\frac{2\vartheta y}{x^{2}+y^{2}}\frac{\nu_{1}}{x}(1-R_{1})
\right]\\
&-\frac{1}{2M}\sigma_{z}\vartheta\,\delta^{(2)}(\mathbf{r})
 +\frac{1}{2}M\omega^{2}(x^{2}+y^{2}).
\end{aligned}
\label{eq:hamiltonian_full}
\end{equation}

\subsection{Separation in Polar Coordinates}

Transforming to polar coordinates $(r,\varphi)$ with
$x=r\cos\varphi$, $y=r\sin\varphi$, the Hamiltonian separates as
\begin{equation}
\begin{aligned}
H={}&-\frac{1}{2M}\frac{\partial^{2}}{\partial r^{2}}
     -\frac{1+2\nu_{1}+2\nu_{2}}{2Mr}\frac{\partial}{\partial r}
     +\frac{1}{2}M\omega^{2}r^{2}\\
&+\frac{1}{2Mr^{2}}\!\left(
  2\mathcal{B}_{\varphi}
 -2\vartheta\mathcal{J}_{\varphi}
 +2\vartheta\sigma_{z}(\nu_{1}R_{1}+\nu_{2}R_{2})
\right)
-\frac{\vartheta}{2M}\delta^{(2)}(\mathbf{r})\sigma_{z},
\end{aligned}
\label{eq:hamiltonian_polar}
\end{equation}
where the coefficient $1+2\nu_{1}+2\nu_{2}$ of the radial derivative
reflects the effective dimension induced by the Dunkl deformation.

In polar coordinates the reflection operators act as
\begin{equation}
R_{1}\Phi(r,\varphi)=\Phi(r,\pi-\varphi),\qquad
R_{2}\Phi(r,\varphi)=\Phi(r,-\varphi),
\label{eq:reflections_polar}
\end{equation}
consistent with the Cartesian definitions
$R_{1}f(x,y)=f(-x,y)$ and $R_{2}f(x,y)=f(x,-y)$.

The Dunkl angular operators $\mathcal{B}_{\varphi}$ and
$\mathcal{J}_{\varphi}$ are defined as \cite{Genest2013a,Genest2013b}
\begin{align}
\mathcal{B}_{\varphi}&=-\frac{1}{2}\frac{\partial^{2}}{\partial\varphi^{2}}
+(\nu_{1}\tan\varphi-\nu_{2}\cot\varphi)\frac{\partial}{\partial\varphi}
+\frac{\nu_{1}}{2\cos^{2}\!\varphi}(1-R_{1})
+\frac{\nu_{2}}{2\sin^{2}\!\varphi}(1-R_{2}),
\label{eq:Bphi}\\
\mathcal{J}_{\varphi}&=i\!\left[
  \frac{\partial}{\partial\varphi}
 +\nu_{2}\cot\varphi(1-R_{2})
 -\nu_{1}\tan\varphi(1-R_{1})
\right],
\label{eq:Jphi}
\end{align}
and satisfy the quadratic relation
\begin{equation}
\mathcal{J}_{\varphi}^{2}=2\mathcal{B}_{\varphi}
+2\nu_{1}\nu_{2}(1-R_{1}R_{2}).
\label{eq:J2B}
\end{equation}

The AB vector potential $e\mathbf{A}=-\vartheta r^{-1}\mathbf{u}_{\varphi}$
depends only on $r$ and satisfies
$R_{j}(e\mathbf{A})=-e\mathbf{A}$ under both reflections
\eqref{eq:reflections_polar}, since the azimuthal unit vector
$\mathbf{u}_{\varphi}$ transforms as
$\mathbf{u}_{\varphi}\mapsto-\mathbf{u}_{\varphi}$ for both $R_{1}$
and $R_{2}$.  Consequently, the Hamiltonian commutes with $R_{1}$
and $R_{2}$, the reflection sectors labelled by $(\epsilon_{1},\epsilon_{2})$
are preserved, and the boundary conditions at the regularised flux
tube are compatible with the reflection symmetry.

\subsection{Angular Eigenvalue Problem}

Since $\mathcal{J}_{\varphi}$, $R_{1}$, and $R_{2}$ commute within
each parity sector, we seek factorised solutions
\begin{equation}
\psi(r,\varphi)=R(r)\,\Phi_{\epsilon}(\varphi)\,\chi_{m_{s}},
\label{eq:factorized}
\end{equation}
where $\chi_{m_{s}}$ are spin eigenfunctions satisfying
$\sigma_{z}\chi_{m_{s}}=m_{s}\chi_{m_{s}}$ with $m_{s}=\pm1$, and
$\Phi_{\epsilon}(\varphi)$ are eigenfunctions of~$\mathcal{J}_{\varphi}$:
\begin{equation}
\mathcal{J}_{\varphi}\Phi_{\epsilon}(\varphi)
=\lambda_{\epsilon}\Phi_{\epsilon}(\varphi).
\label{eq:Jeigen}
\end{equation}
The subscript $\epsilon=\epsilon_{1}\epsilon_{2}$ labels the joint
reflection eigenvalues, with $R_{j}\Phi_{\epsilon}=\epsilon_{j}\Phi_{\epsilon}$
and $\epsilon_{j}=\pm1$.  Using \eqref{eq:J2B}, the eigenvalues of
$\mathcal{B}_{\varphi}$ are expressed in terms of~$\lambda_{\epsilon}$,
allowing the angular dependence to be fully encoded in the radial
equation.

\paragraph{Even sector $\epsilon=+1$ ($\epsilon_{1}=\epsilon_{2}=\pm1$):}
\begin{align}
\Phi_{+}(\varphi)&=A_{l}P_{l}^{(\nu_{1}-\frac{1}{2},\nu_{2}-\frac{1}{2})}
(\cos2\varphi)
\pm iA'_{l}\sin\varphi\cos\varphi\,
P_{l-1}^{(\nu_{1}+\frac{1}{2},\nu_{2}+\frac{1}{2})}(\cos2\varphi),\\
\lambda_{+}&=\pm2\sqrt{l(l+\nu_{1}+\nu_{2})},\qquad l\in\mathbb{N}^{*}.
\end{align}

\paragraph{Odd sector $\epsilon=-1$ ($\epsilon_{1}=-\epsilon_{2}$):}

Let $l=n+\tfrac{1}{2}$ with $n\in\mathbb{N}$.  Then
\begin{align}
\Phi_{-}(\varphi)&=B_{n}\cos\varphi\,
P_{n}^{(\nu_{1}+\frac{1}{2},\nu_{2}-\frac{1}{2})}(\cos2\varphi)
\mp iB'_{n}\sin\varphi\,
P_{n}^{(\nu_{1}-\frac{1}{2},\nu_{2}+\frac{1}{2})}(\cos2\varphi),\\
\lambda_{-}&=\pm2\sqrt{(n+\nu_{1}+\tfrac{1}{2})(n+\nu_{2}+\tfrac{1}{2})},
\end{align}
where $P_{n}^{(a,b)}$ denote Jacobi polynomials and the
normalisation constants are expressed in terms of Gamma functions in
the standard way.

\subsection{Radial Equation and Regularisation}

Using $\sigma_{z}\chi_{m_{s}}=m_{s}\chi_{m_{s}}$ and the polar
form $\delta^{(2)}(\mathbf{r})=\tfrac{1}{r}\delta(r)$, the singular
magnetic term reduces to $-\tfrac{\vartheta}{2M}m_{s}\tfrac{\delta(r)}{r}$.
Inserting the angular eigenfunctions into \eqref{eq:hamiltonian_polar}
and substituting $\mathcal{Q}(r)=r^{-\delta}\mathcal{L}(r)$ with
$\delta=\tfrac{1}{2}+\nu_{1}+\nu_{2}$, we obtain the radial equation
\begin{equation}
\left[
  -\frac{\mathrm{d}^{2}}{\mathrm{d}r^{2}}
  +\frac{(\vartheta-\lambda_{\epsilon})^{2}
         +\delta(\delta-1)-2\nu_{1}\nu_{2}(1-\epsilon)
         +2\vartheta(\nu_{1}\epsilon_{1}+\nu_{2}\epsilon_{2})m_{s}}{r^{2}}
  -\vartheta m_{s}\frac{\delta(r)}{r}
  +M^{2}\omega^{2}r^{2}
\right]\mathcal{L}(r)=2ME\,\mathcal{L}(r).
\label{eq:radial_raw}
\end{equation}
The distributional term $\delta(r)/r$ renders the operator singular
at the origin.  Following the self-adjoint extension approach, we
regularise by replacing the zero-radius flux tube with a finite one
of radius~$r_{0}$:
\begin{equation}
\frac{\delta(r)}{r}\longrightarrow\frac{\delta(r-r_{0})}{r},\qquad
e\mathbf{A}=-\frac{\vartheta}{r}\theta(r-r_{0})\mathbf{u}_{\varphi},
\label{eq:regularization}
\end{equation}
and we take the limit $r_{0}\to0^{+}$ at the end.

Using the algebraic identity
\begin{equation}
\delta(\delta-1)-2\nu_{1}\nu_{2}(1-\epsilon)
=(\nu_{1}+\epsilon\nu_{2})^{2}-\tfrac{1}{4},
\label{eq:identity}
\end{equation}
we define the effective angular momenta for the inner and outer regions:
\begin{align}
K_{\text{in}}^{2}&=\lambda_{\epsilon}^{2}+(\nu_{1}+\epsilon\nu_{2})^{2},
\label{eq:K_in_def}\\
K_{\text{out}}^{2}&=(\vartheta-\lambda_{\epsilon})^{2}
           +(\nu_{1}+\epsilon\nu_{2})^{2}
           +2\vartheta(\nu_{1}\epsilon_{1}+\nu_{2}\epsilon_{2})m_{s}.
\label{eq:K_out_def}
\end{align}
The radial equations in the inner ($r<r_{0}$) and outer ($r>r_{0}$)
regions then take the standard form
\begin{align}
\left[
  \frac{\mathrm{d}^{2}}{\mathrm{d}r^{2}}
  -\frac{K_{\text{in}}^{2}-\tfrac{1}{4}}{r^{2}}
  -M^{2}\omega^{2}r^{2}+2ME_{\text{in}}
\right]\mathcal{L}_{\text{in}}(r)&=0,\quad r<r_{0},
\label{eq:radial_in}\\
\left[
  \frac{\mathrm{d}^{2}}{\mathrm{d}r^{2}}
  -\frac{K_{\text{out}}^{2}-\tfrac{1}{4}}{r^{2}}
  -M^{2}\omega^{2}r^{2}+2ME_{\text{out}}
\right]\mathcal{L}_{\text{out}}(r)&=0,\quad r>r_{0}.
\label{eq:radial_out}
\end{align}

\subsection{Hilbert-Space Structure, Square Integrability,
            and Self-Adjointness}
\label{sec:hilbert}

The natural Hilbert space associated with the Dunkl Laplacian is
$\mathcal{H}=L^{2}(\mathbb{R}^{2}\setminus\{0\},d\mu_{D})\otimes\mathbb{C}^{2}$,
with Dunkl measure
\begin{equation}
d\mu_{D}=r^{1+2\nu_{1}+2\nu_{2}}\,dr\,d\varphi.
\label{eq:dunkl_measure}
\end{equation}
The inner product is accordingly
\begin{equation}
\langle\Psi,\Phi\rangle_{D}
=\int_{0}^{2\pi}\int_{0}^{\infty}
\Psi^{\dagger}(r,\varphi)\Phi(r,\varphi)\,
r^{1+2\nu_{1}+2\nu_{2}}\,dr\,d\varphi.
\label{eq:dunkl_inner}
\end{equation}

Substitution $\mathcal{Q}(r)=r^{-\delta}\mathcal{L}(r)$ with
$\delta=\tfrac{1}{2}+\nu_{1}+\nu_{2}$ converts the Dunkl norm to
a standard $L^{2}(0,\infty)$ norm:
\begin{equation}
\|\mathcal{Q}\|_{D}^{2}
=\int_{0}^{\infty}|\mathcal{Q}(r)|^{2}\,r^{1+2\nu_{1}+2\nu_{2}}\,dr
=\int_{0}^{\infty}|\mathcal{L}(r)|^{2}\,dr.
\label{eq:Q_transform}
\end{equation}
The reduced radial operator acting on $\mathcal{L}(r)$ is
\begin{equation}
\mathcal{L}_{\mathrm{rad}}
=-\frac{d^{2}}{dr^{2}}
 +\frac{K^{2}-\tfrac{1}{4}}{r^{2}}
 +M^{2}\omega^{2}r^{2},
\label{eq:SL_operator}
\end{equation}
which is a standard Sturm--Liouville operator on $L^{2}(0,\infty)$.
Here $K$ denotes either $K_{\text{in}}$ or $K_{\text{out}}$ depending
on the region.

Near the origin, the normalizable solutions behave as
$\mathcal{L}(r)\sim r^{K+1/2}$, giving
$\|\mathcal{L}\|^{2}\sim\int_{0}r^{2K+1}\,dr$,
which converges if and only if
\begin{equation}
K>-1.
\label{eq:regularity_condition}
\end{equation}
This square-integrability condition determines the admissible angular
sectors and gives rise to the lowest allowed angular quantum
number~$\ell_{0}$.

Regarding self-adjointness: the endpoint $r=\infty$ is always in the
limit-point case because of the confining harmonic term.  At the
origin, the Weyl limit-point/limit-circle criterion applied to the
inverse-square potential $\tfrac{K^{2}-1/4}{r^{2}}$ gives
\begin{itemize}
\item limit-point (essentially self-adjoint) when $|K|\geq1$;
\item limit-circle (one-parameter family of self-adjoint extensions)
      when $|K|<1$.
\end{itemize}
The present work focuses on the self-adjoint realisation selected by
the finite-radius regularisation \eqref{eq:regularization}.  A
complete classification of all self-adjoint extensions and their
associated spectral constraints lies beyond the scope of this work.

\subsection{Radial Solutions and Matching Conditions}
\label{sec:matching}

The radial equations \eqref{eq:radial_in} and \eqref{eq:radial_out}
are singular oscillator equations and admit normalizable solutions in
terms of generalized Laguerre polynomials. Imposing square
integrability at infinity yields
\begin{align}
\mathcal{L}_{\text{in}}(r)&=N_{\text{in}} r^{K_{\text{in}}+\frac{1}{2}}
  e^{-M\omega r^{2}/2}L_{n}^{K_{\text{in}}}(M\omega r^{2}),\quad r<r_{0},
\label{eq:sol_inner}\\
\mathcal{L}_{\text{out}}(r)&=N_{\text{out}} r^{K_{\text{out}}+\frac{1}{2}}
  e^{-M\omega r^{2}/2}L_{n}^{K_{\text{out}}}(M\omega r^{2}),\quad r>r_{0},
\label{eq:sol_outer}
\end{align}
with corresponding energy eigenvalues
\begin{align}
E_{\text{in}}&=\omega(2n+K_{\text{in}}+1),\qquad n=0,1,2,\ldots,\label{eq:E_inner}\\
E_{\text{out}}&=\omega(2n+K_{\text{out}}+1),\qquad n=0,1,2,\ldots.\label{eq:E_outer}
\end{align}
The regularisation procedure follows the treatment introduced by
Hagen \cite{Hagen1990} and applied to Pauli--AB systems by Bouguerra
et al.\ \cite{Bouguerra2008}.  Integrating \eqref{eq:radial_raw}
across $(r_{0}-\eta,r_{0}+\eta)$ and taking $\eta\to0^{+}$ yields
the matching conditions
\begin{align}
\mathcal{L}_{\text{in}}(r_{0})&=\mathcal{L}_{\text{out}}(r_{0}),
\label{eq:match_cont}\\
\left.\frac{d\mathcal{L}_{\text{out}}}{dr}\right|_{r_{0}}
-\left.\frac{d\mathcal{L}_{\text{in}}}{dr}\right|_{r_{0}}
&=-\frac{\vartheta m_{s}}{r_{0}}\,\mathcal{L}_{\text{in}}(r_{0}).
\label{eq:match_jump}
\end{align}

In the limit $r_{0}\to0^{+}$, using
$L_{n}^{\alpha}(x)=1+\mathcal{O}(x)$ as $x\to0$, the dominant
behaviour of the solutions near the origin is
\begin{equation}
\mathcal{L}_{\text{in}}(r)\sim N_{\text{in}}r^{K_{\text{in}}+\frac{1}{2}},\qquad
\frac{d\mathcal{L}_{\text{in}}}{dr}\sim N_{\text{in}}
\!\left(K_{\text{in}}+\tfrac{1}{2}\right)r^{K_{\text{in}}-\frac{1}{2}},
\label{eq:expansion_in}
\end{equation}
and
\begin{equation}
\mathcal{L}_{\text{out}}(r)\sim N_{\text{out}}r^{K_{\text{out}}+\frac{1}{2}},\qquad
\frac{d\mathcal{L}_{\text{out}}}{dr}\sim N_{\text{out}}
\!\left(K_{\text{out}}+\tfrac{1}{2}\right)r^{K_{\text{out}}-\frac{1}{2}}.
\label{eq:expansion_out}
\end{equation}
Continuity \eqref{eq:match_cont} gives
$N_{\text{out}}=N_{\text{in}}r_{0}^{K_{\text{in}}-K_{\text{out}}}$.
Substituting into the jump condition \eqref{eq:match_jump} and
dividing by $N_{\text{in}}r_{0}^{K_{\text{in}}-1/2}$ yields, to
leading order,
\begin{equation}
K_{\text{out}}=K_{\text{in}}-\vartheta m_{s}.
\label{eq:Krelation}
\end{equation}

To derive the compatibility condition, we equate the two expressions
for~$K_{\text{out}}^{2}$.  From \eqref{eq:K_out_def}, using
$\nu_{1}\epsilon_{1}+\nu_{2}\epsilon_{2}
=\nu_{1}+\varepsilon\nu_{2}$,
\begin{equation}
K_{\text{out}}^{2}=(\vartheta-\lambda_{\varepsilon})^{2}
+(\nu_{1}+\varepsilon\nu_{2})^{2}
+2\vartheta m_{s}(\nu_{1}+\varepsilon\nu_{2}).
\end{equation}
Squaring \eqref{eq:Krelation} and substituting
$K_{\text{in}}^{2}=\lambda_{\varepsilon}^{2}
+(\nu_{1}+\varepsilon\nu_{2})^{2}$,
equating the two expressions for $K_{\text{out}}^{2}$ and simplifying
gives
\begin{equation}
(\nu_{1}+\varepsilon\nu_{2})\bigl(K_{\text{in}}-m_{s}\lambda_{\varepsilon}\bigr)=0.
\label{eq:consistency_general}
\end{equation}
For nontrivial angular sectors ($\lambda_{\varepsilon}\neq0$), a
direct check shows that the second factor cannot vanish unless
$\nu_{1}+\varepsilon\nu_{2}=0$ as well.  The finite-radius
regularisation therefore selects the unique admissible condition
\begin{equation}
\nu_{1}+\varepsilon\nu_{2}=0.
\label{eq:symconstraint}
\end{equation}
This condition is a consequence of the specific self-adjoint
realisation selected by the regularisation scheme
\eqref{eq:regularization}; other choices of boundary conditions at
the origin may yield different constraints.  Its physical content is
a compatibility relation between the discrete reflection parity
$\varepsilon=\pm1$ and the Dunkl deformation parameters: the singular
AB topology correlates these two structures, restricting the
admissible reflection sectors.  In the absence of the AB flux
($\vartheta=0$), condition \eqref{eq:symconstraint} disappears and
$\nu_{1}$, $\nu_{2}$ remain independent, consistently recovering the
standard Dunkl oscillator.

With condition \eqref{eq:symconstraint} imposed, the effective
angular momenta simplify to
$K_{\text{in}}=|\lambda_{\varepsilon}|/m_{s}$ and
$K_{\text{out}}=K_{\text{in}}-\vartheta m_{s}$,
and the physical energy spectrum (outer region) is
\begin{equation}
E_{n,l,m_{s}}^{(\text{out})}
=\omega\bigl(2n+|\lambda_{\epsilon}|-\vartheta m_{s}+1\bigr),\qquad
n=0,1,2,\ldots.
\label{eq:spectrum_final}
\end{equation}
These energy levels define the spectrum used in all subsequent
thermodynamic analysis.

\section{Thermodynamic Properties}
\label{sec:thermo}

\subsection{Partition Function}
\label{sec:partition}

The canonical partition function is
\begin{equation}
Z(\beta)=\sum_{n,\ell,m_{s}}e^{-\beta E_{n,\ell,m_{s}}},\qquad
\beta=\frac{1}{k_{B}T}.
\label{eq:Z_def}
\end{equation}
The intermediate steps leading to the compact form below are given
explicitly in Appendix~\ref{app:partition}.  The result is
\begin{equation}
Z(\beta)=\frac{2\,e^{-\beta E_{0}}\cosh(\beta\omega\vartheta)}
              {\bigl(1-e^{-2\beta\omega}\bigr)^{2}},
\label{eq:Zgeneral}
\end{equation}
where the ground-state energy is
\begin{equation}
E_{0}=
\begin{cases}
\omega(1+2\ell_{0}), & \varepsilon=+1,\\[4pt]
\omega\bigl(1+2(\ell_{0}+\nu)\bigr), & \varepsilon=-1.
\end{cases}
\label{eq:E0cases}
\end{equation}
The quantity~$\ell_{0}$ is the lowest admissible angular quantum
number, determined by the square-integrability condition
$K_{\text{out}}>-1$ established in Sec.~\ref{sec:hilbert}.

\subsubsection*{Determination of $\ell_{0}$}

The regularity condition for the outer region reads
$K_{\text{out}}>-1$. Using the matching condition
$K_{\text{out}}=K_{\text{in}}-\vartheta m_{s}$, we obtain
$K_{\text{in}}-\vartheta m_{s}>-1$.
The most restrictive case corresponds to the spin orientation that
maximises the flux contribution.

\paragraph{Even sector ($\varepsilon=+1$).}

The compatibility condition $\nu_{1}=-\nu_{2}$ gives
$\lambda_{+}=\pm2\ell$ with $\ell\in\mathbb{N}^{*}$, so
$K_{\text{in}}=2\ell$ and the regularity condition becomes
\begin{equation}
2\ell-\vartheta m_{s}>-1.
\label{eq:even_condition}
\end{equation}
The smallest allowed value in this sector is $\ell=1$.  Substituting
into \eqref{eq:even_condition} gives $\vartheta m_{s}<3$, so the
threshold is $\vartheta m_{s}=3$.  More precisely, the condition
$\lceil\tfrac{\vartheta m_{s}-1}{2}\rceil\leq1$ is equivalent to
$\vartheta m_{s}\leq3$.  Hence
\begin{equation}
\ell_{0}=
\begin{cases}
1, & \vartheta m_{s}\leq3,\\[4pt]
\left\lceil\dfrac{\vartheta m_{s}-1}{2}\right\rceil,
& \vartheta m_{s}>3.
\end{cases}
\label{eq:l0_even_final}
\end{equation}

\paragraph{Odd sector ($\varepsilon=-1$).}

The compatibility condition $\nu_{1}=\nu_{2}\equiv\nu$ gives
$\lambda_{-}=\pm2(\ell+\nu)$ with $\ell=n+\tfrac{1}{2}$,
$n\in\mathbb{N}$, so $K_{\text{in}}=2(\ell+\nu)$ and the regularity
condition becomes
\begin{equation}
2(\ell+\nu)-\vartheta m_{s}>-1.
\label{eq:odd_condition}
\end{equation}
The smallest allowed value is $\ell=\tfrac{1}{2}$.  Substituting
gives $\vartheta m_{s}<2(1+\nu)$, so the threshold is
$\vartheta m_{s}=2(1+\nu)$, which follows from
$\lceil\tfrac{\vartheta m_{s}-2\nu-1}{2}\rceil\leq0$.  Hence
\begin{equation}
\ell_{0}=
\begin{cases}
\tfrac{1}{2}, & \vartheta m_{s}\leq2(1+\nu),\\[4pt]
\left\lceil\dfrac{\vartheta m_{s}-2\nu-1}{2}\right\rceil
+\tfrac{1}{2}, & \vartheta m_{s}>2(1+\nu).
\end{cases}
\label{eq:l0_odd_final}
\end{equation}

\medskip

\noindent
The boundary values are included in the protected regime by
convention.  The critical case corresponds to the integrability
threshold $K_{\text{out}}=-1$, where the radial wavefunction remains
square-integrable; the inequality $K_{\text{out}}>-1$ is therefore
strict, but including the boundary point in the protected regime is a
matter of convention that does not affect the physical content.

In the even sector, the thermodynamic quantities are independent of the specific magnitudes of $\nu_1$ and $\nu_2$; they depend only on the compatibility relation $\nu_1=-\nu_2$.  Any choice satisfying $-\frac{1}{2}<\nu_1<\frac{1}{2}$ yields the same curves.

\subsubsection*{Physical interpretation of $\ell_{0}$}

In the conventional Pauli--AB problem, the magnetic flux shifts the
angular spectrum but excludes no sector: every angular quantum number
remains admissible.  In the Dunkl-deformed setting the situation is
qualitatively different.  The square-integrability condition
$K_{\text{out}}>-1$, defined with respect to the Dunkl measure
\eqref{eq:dunkl_measure}, converts the flux-induced shift into a
genuine state-exclusion mechanism.  As~$\vartheta$ increases,
$K_{\text{out}}$ may fall below the regularity threshold for the
lowest Dunkl-admissible sector, at which point the corresponding
state ceases to be normalizable and must be removed from the physical
spectrum.  The lowest admissible quantum number~$\ell_{0}$ is
therefore the smallest value of~$\ell$ for which $K_{\text{out}}>-1$
is restored.

This mechanism gives rise to the crossover values in
\eqref{eq:l0_even_final} and \eqref{eq:l0_odd_final}.  Below the
threshold, $\ell_{0}$ remains at its flux-free value and the Dunkl
parity constraint alone determines the ground-state structure.
Above threshold, the AB-induced regularity condition dominates
and~$\ell_{0}$ acquires an explicit flux dependence.  In the odd
sector, the threshold $2(1+\nu)$ depends directly on~$\nu$: a larger
Dunkl parameter enlarges the flux interval over which the lowest
angular sector remains admissible, demonstrating that~$\nu$ controls
both the energy scale and the robustness of the ground state against
the topological perturbation.

This dual role of~$\nu$ is reflected in the thermodynamic
quantities.  The partition function \eqref{eq:Zgeneral} and the
internal energy (Sec.~\ref{sec:U}) depend explicitly on~$\ell_{0}$
and~$\nu$ through~$E_{0}$.  By contrast, the entropy and heat
capacity are obtained via combinations---$S=\beta(U-F)$ and
$C_{V}=\partial U/\partial T$---in which the additive
temperature-independent contribution~$E_{0}$ cancels exactly.
Consequently, within a fixed $\ell_{0}$ branch, $S$ and $C_{V}$ are
independent of both~$\ell_{0}$ and~$\nu$; across branches,~$\nu$
and~$\ell_{0}$ determine which angular sectors are admissible, and
thus indirectly affect all thermodynamic quantities through the
branch selection.

\begin{figure}[H]
\centering
\includegraphics[width=0.4\textwidth]{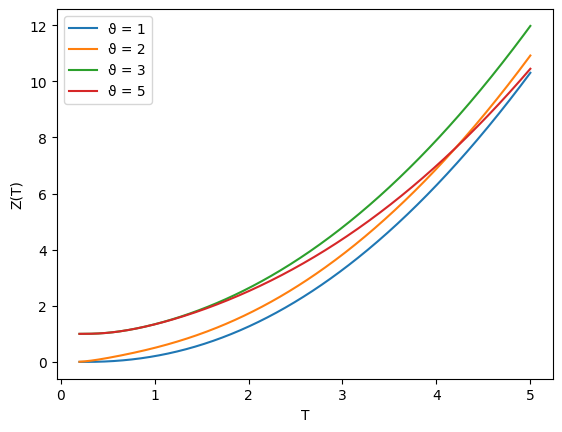}
\caption{Temperature dependence of $Z(T)$ for $\varepsilon=+1$ and
         several values of the AB flux~$\vartheta$.  The separation
         between curves at low temperature reflects the
         flux-dependent shift of~$\ell_{0}$; the curves converge
         at high temperature as thermal excitations dominate.
         }
\label{fig:partition+}
\end{figure}

\begin{figure}[H]
\centering
\begin{subfigure}{0.4\textwidth}
\centering
\includegraphics[width=\linewidth]{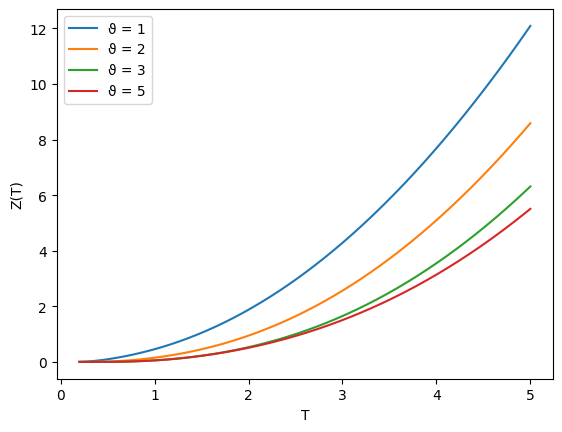}
\caption{$\nu=-0.4$}
\end{subfigure}
\hfill
\begin{subfigure}{0.4\textwidth}
\centering
\includegraphics[width=\linewidth]{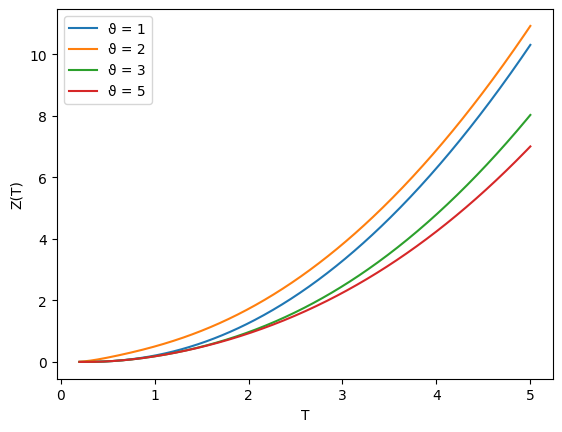}
\caption{$\nu=0.0$}
\end{subfigure}

\vspace{0.4cm}

\begin{subfigure}{0.4\textwidth}
\centering
\includegraphics[width=\linewidth]{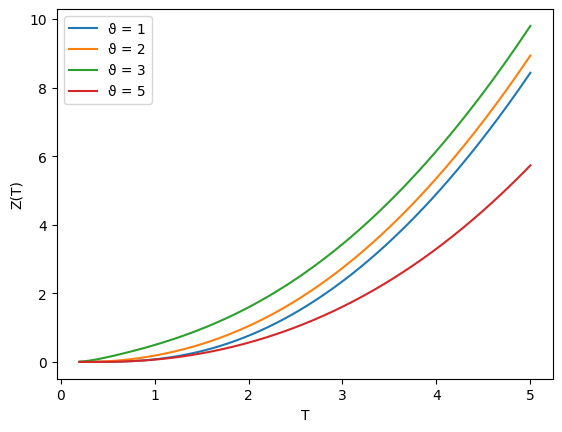}
\caption{$\nu=0.5$}
\end{subfigure}
\hfill
\begin{subfigure}{0.4\textwidth}
\centering
\includegraphics[width=\linewidth]{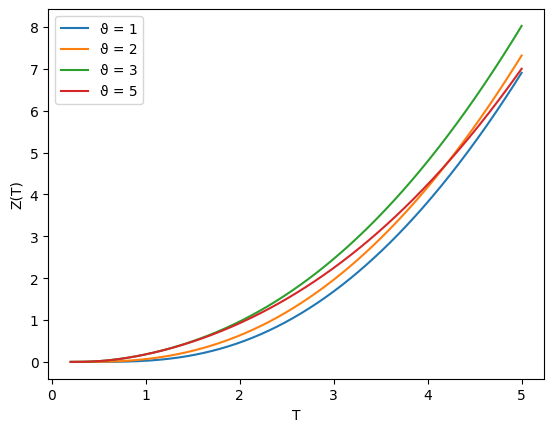}
\caption{$\nu=1.0$}
\end{subfigure}

\caption{Temperature dependence of $Z(T)$ for $\varepsilon=-1$ and
         several values of the Dunkl parameter~$\nu$ and the
         AB flux~$\vartheta$.  The additional dependence on~$\nu$,
         absent in the even sector, originates from the
         deformation-induced shift in the ground-state energy and
         from the $\nu$-dependent threshold of \eqref{eq:l0_odd_final}.}
\label{fig:Zminus}
\end{figure}

To illustrate these effects, Figs.~\ref{fig:partition+} and
\ref{fig:Zminus} display the temperature dependence of the partition
function in the even and odd sectors, respectively. 

These curves are drawn for $\nu_1=-\nu_2$, as required
         by the even-sector compatibility relation; per the discussion
         above, any choice satisfying $-\tfrac12<\nu_1<\tfrac12$ yields
         identical curves.

The additional dependence on~$\nu$,
         absent in the even sector, originates from the
         deformation-induced shift in the ground-state energy and
         from the $\nu$-dependent threshold of \eqref{eq:l0_odd_final}.
         These plots use $\nu_1=\nu_2=\nu$ with the explicit values
         shown in each subcaption.
\subsection{Internal Energy}
\label{sec:U}

From $U=-\partial_{\beta}\ln Z$, one obtains
\begin{equation}
U(\beta)=E_{0}-\omega\vartheta\tanh(\beta\omega\vartheta)
+\frac{4\omega}{e^{2\beta\omega}-1}.
\label{eq:Ugeneral}
\end{equation}
At low temperatures $U\to E_{0}-\omega\vartheta\,\mathrm{sgn}(\vartheta)$,
showing that the AB flux induces a nontrivial shift that depends on
the sign of~$\vartheta$.  As the temperature increases, thermal
excitations populate higher oscillator states and $U(T)$ grows
toward the classical limit $U\approx2k_{B}T$.

\begin{figure}[H]
\centering
\includegraphics[width=0.48\textwidth]{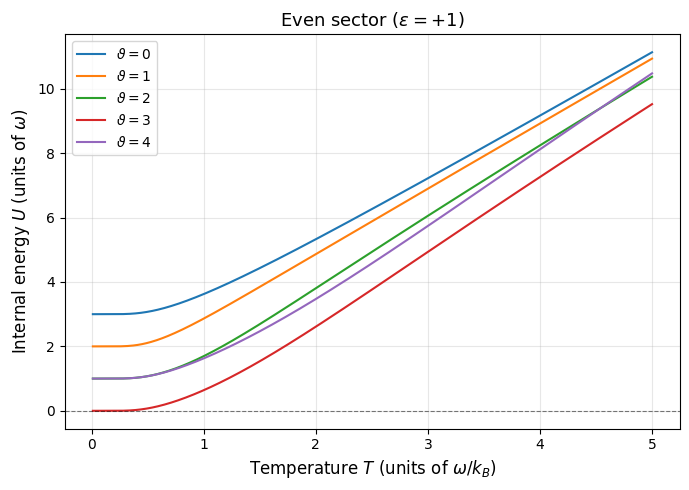}
\caption{Temperature dependence of $U(T)$ for $\varepsilon=+1$ and
         several values of~$\vartheta$.}
\label{fig:Inter Energy eps plus}
\end{figure}

\begin{figure}[H]
\centering
\includegraphics[width=0.9\textwidth]{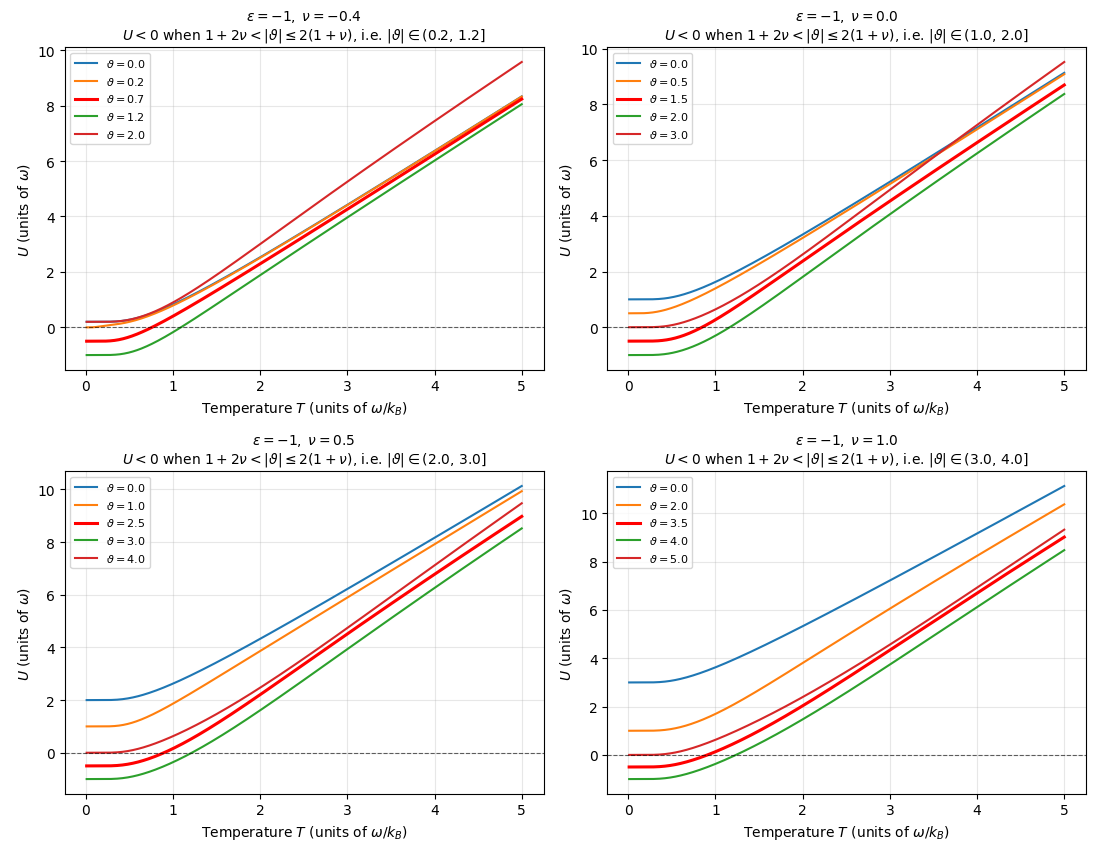}
\caption{Temperature dependence of $U(T)$ for $\varepsilon=-1$ and
         several values of~$\nu$ and~$\vartheta$. These plots
         use $\nu_1=\nu_2=\nu$ with the values $\nu=-0.4,0,0.5,1.0$
         as indicated.}
\label{fig:Inter Energy eps minus}
\end{figure}

In the even sector (Fig.~\ref{fig:Inter Energy eps plus}), the
curves for different~$\vartheta$ separate at intermediate
temperatures and converge at high temperatures, where thermal
fluctuations dominate.  In the odd sector
(Fig.~\ref{fig:Inter Energy eps minus}), the Dunkl parameter~$\nu$
introduces an additional shift through the modified ground-state
energy $E_{0}=\omega(1+2(\ell_{0}+\nu))$; larger values of~$\nu$
lead to higher ground-state energies and more pronounced thermal
activation.

As in
         Fig.~\ref{fig:partition+}, these curves are independent of
         the specific magnitudes of $\nu_1$ and $\nu_2$; only the
         compatibility relation $\nu_1=-\nu_2$ matters.

\subsection{Entropy}
\label{sec:S}

From $S=\beta(U-F)$,
\begin{equation}
S(\beta)=\ln\!\bigl[2\cosh(\beta\omega\vartheta)\bigr]
-2\ln\!\bigl(1-e^{-2\beta\omega}\bigr)
-\beta\omega\vartheta\tanh(\beta\omega\vartheta)
+\frac{4\beta\omega}{e^{2\beta\omega}-1}.
\label{eq:Sgeneral}
\end{equation}
The entropy increases monotonically with temperature and satisfies
$S\to0$ as $T\to0$ (third law), since the system occupies a unique
ground state determined by~$\ell_{0}$.  Within a fixed branch,
$S$ is independent of~$\ell_{0}$ and~$\nu$, because~$E_{0}$ cancels
identically in $S=\beta(U-F)$.  However,~$\nu$ determines which
branch the system occupies and thereby controls the accessible state
counting across branches.

\begin{figure}[H]
\centering
\includegraphics[width=0.5\textwidth]{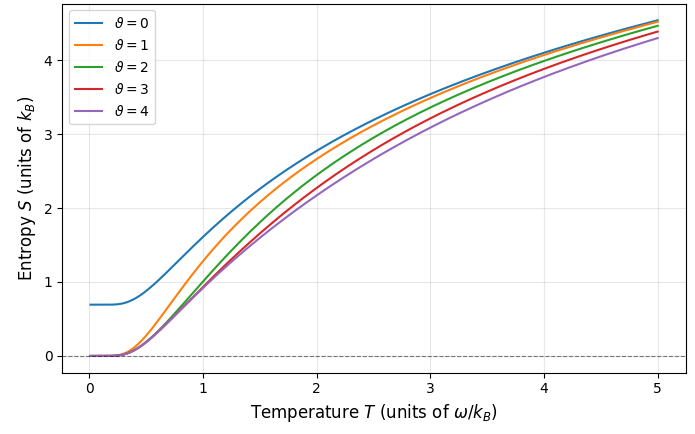}
\caption{Temperature dependence of $S(T)$ for several values
         of~$\vartheta$.}
\label{fig:Entropy}
\end{figure}

The AB flux delays the onset of entropy growth by increasing the
effective ground-state energy gap (Fig.~\ref{fig:Entropy}).

Since $S$ is independent of both
         $\ell_0$ and $\nu$ within a fixed branch, the values of
         $\nu_1,\nu_2$ are not required to generate these universal
         curves.
\subsection{Heat Capacity}
\label{sec:CV}

From $C_{V}=\partial U/\partial T=-\beta^{2}\partial_{\beta}U$,
\begin{equation}
C_{V}(\beta)=\beta^{2}\omega^{2}\!\left[
  \vartheta^{2}\mathrm{sech}^{2}(\beta\omega\vartheta)
 +2\,\mathrm{csch}^{2}(\beta\omega)
\right].
\label{eq:Cgeneral}
\end{equation}
The heat capacity exhibits a pronounced Schottky-like peak, which is
a direct signature of the finite energy gap introduced by the AB
flux.  We use the term ``Schottky-\emph{like}'' deliberately: unlike
a textbook Schottky anomaly, where $C_{V}\to0$ at high temperature
once a finite number of levels saturates, here $C_{V}\to2k_{B}$ as
$T\to\infty$, reflecting the unbounded oscillator ladder.  The
resemblance to the Schottky profile is therefore confined to the
low- and intermediate-temperature peak structure.

\begin{figure}[H]
\centering
\includegraphics[width=0.5\textwidth]{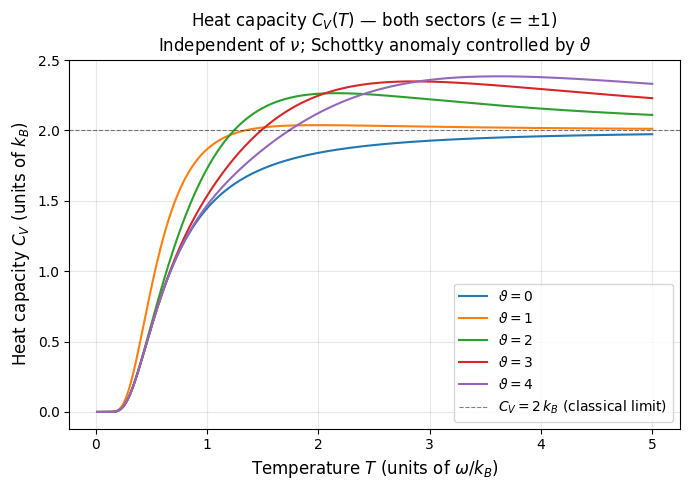}
\caption{Heat capacity $C_{V}(T)$ for several values of~$\vartheta$.
         The dashed line marks the classical equipartition
         value $2k_{B}$.}
\label{fig:CV}
\end{figure}

The peak position shifts to higher temperatures as~$\vartheta$
increases, confirming that the flux enhances the characteristic
excitation energy.  At low temperatures, $C_{V}\to0$ as expected.
The heat capacity is independent of~$\ell_{0}$ and~$\nu$ within a
fixed branch, since~$E_{0}$ is removed by differentiation; however,
across branches, these parameters determine which angular sectors are
admissible and thereby affect $C_{V}$ indirectly through branch
selection (Fig.~\ref{fig:CV}).

These curves are universal and
         independent of the specific values of $\nu_1,\nu_2$, since
         $E_0$ cancels in the derivative defining $C_V$.
\section{Towards Physical Realisations of the Dunkl Deformation}
\label{sec:realization}

The results above raise a natural question: in what physical setting
could the Dunkl reflection parameters $\nu_{1},\nu_{2}$---and hence
the compatibility condition \eqref{eq:symconstraint} and the
constraint on~$\ell_{0}$---actually arise?  Two routes appear most
promising.

The first is an algebraic route: instead of seeking a literal
discrete spatial reflection symmetry, one can engineer an effective
Dunkl-type kinetic term in synthetic platforms where the reflection
operators are realised through an engineered discrete symmetry of a
driving protocol.  A precedent is found in generalised Rabi-type
models, where driven trapped-ion and circuit-QED platforms have been
shown to realise effective Hamiltonians whose spin-subspace
diagonalisation is governed by differential operators of Dunkl type,
with the relevant reflection symmetry originating from an engineered
discrete ($\mathbb{Z}_{2}$ or higher) symmetry of the drive.  An
analogous strategy---engineering a discrete symmetry in the AB-ring
geometry through patterned gating, periodic driving, or a synthetic
lattice with a built-in parity symmetry---could in principle realise
an effective Dunkl--Pauli Hamiltonian of the type studied here.

The second, more direct route is to interpret the Dunkl deformation
phenomenologically, as an effective description of a ring with a
sharply localised, parity-symmetric scattering potential at the
solenoid position---for instance, a point defect or an engineered
barrier respecting the reflection symmetry of the ring
geometry---rather than as arising from a Calogero-type interaction.
In this reading, $\nu_{1},\nu_{2}$ parametrise the strength of this
localised perturbation, and the predictions of this work---the
$\ell_{0}$-controlled offset in $Z(T)$ and $U(T)$, and the
Schottky-like peak in $C_{V}(T)$---constitute falsifiable signatures
distinguishing such a system from an ordinary AB--Pauli ring.

We emphasise that a concrete experimental embodiment of the
Dunkl--Pauli Hamiltonian in a solid-state mesoscopic ring remains to
be identified.  Nevertheless, the thermodynamic signatures derived
in Sec.~\ref{sec:thermo} provide concrete, falsifiable targets that
any such realisation should reproduce.

\section{Conclusion}
\label{sec:conclusion}

We have analysed a two-dimensional Dunkl--Pauli oscillator in the
presence of an Aharonov--Bohm flux, focusing on the interplay
between a singular topological gauge field and reflection symmetry
encoded through Dunkl operators.

The main result is the emergence of a \textbf{lowest admissible
angular quantum number}~$\ell_{0}$, arising from the regularity
condition $K_{\text{out}}>-1$ together with the matching conditions at the
flux tube.  This leads to the compatibility relation
$\nu_{1}+\varepsilon\nu_{2}=0$ between the Dunkl parameters and the
reflection sector, which is a consequence of the self-adjoint
realisation selected by the finite-radius regularisation.  The set
of allowed angular momentum states is restricted in a flux- and
deformation-dependent way, resulting in a genuine restriction of the
admissible angular sectors rather than a mere spectral shift.

The quantity~$\ell_{0}$ is not a technical cutoff but a physical
signature of the competition between a parity-imposed angular floor
and an AB-induced regularity exclusion mechanism.  The threshold
values $\vartheta m_{s}\leq3$ (even sector) and
$\vartheta m_{s}\leq2(1+\nu)$ (odd sector) separate a regime where
the Dunkl parity constraint dominates from one where the AB-induced
condition takes over.

The odd sector reveals a particularly rich structure: $\nu$ plays a
dual role.  Within a fixed $\ell_{0}$ branch it contributes an
additive shift to~$E_{0}$; globally it modifies the admissible
angular sectors by shifting the threshold values.  This is reflected
in the thermodynamics: the partition function and internal energy
depend explicitly on~$\ell_{0}$ and~$\nu$ through~$E_{0}$, while
the entropy and heat capacity are independent of both within a fixed
branch (since~$E_{0}$ cancels), but depend indirectly on~$\nu$
through branch selection.

The heat capacity exhibits a flux-controlled Schottky-like peak whose
position is tunable by~$\vartheta$, while~$\nu$ affects the energy
scale without modifying the excitation structure within a branch.
At high temperatures, the system approaches the classical
two-dimensional oscillator limit.

\textbf{These results demonstrate that the Aharonov--Bohm flux,
when combined with Dunkl reflection symmetry, restricts the
admissible angular sectors rather than merely shifting the energy
spectrum, with the emergence of~$\ell_{0}$ as its key manifestation.}

We have also identified two possible routes through which the Dunkl
deformation could in principle be approached experimentally
(Sec.~\ref{sec:realization}), while noting that a concrete
solid-state embodiment remains an open question.  Future extensions
may include time-dependent fluxes, non-Abelian gauge fields, and
higher-dimensional reflection groups.

\appendix
\section{Derivation of the Partition Function}
\label{app:partition}

This appendix gives the intermediate steps connecting the energy spectrum \eqref{eq:spectrum_final} to the compact partition function \eqref{eq:Zgeneral}.

Using the compatibility condition \eqref{eq:symconstraint}, the spectrum takes the explicit form
\begin{equation}
E_{n,\ell,m_{s}}=
\begin{cases}
\omega(2n+2\ell-\vartheta m_{s}+1), & \varepsilon=+1,\\[4pt]
\omega\bigl(2n+2(\ell+\nu)-\vartheta m_{s}+1\bigr), & \varepsilon=-1,
\end{cases}
\label{eq:app_spectrum}
\end{equation}
with $n=0,1,2,\ldots$ and $\ell=\ell_{0},\ell_{0}+1,\ell_{0}+2,\ldots$. Setting $x=e^{-2\beta\omega}$ (with $0<x<1$ for $\beta>0$), both $n$ and $\ell$ enter the exponent additively with the same coefficient~$2\omega$, so the double sum factorises:
\begin{equation}
\sum_{n=0}^{\infty}x^{n}=\frac{1}{1-x},\qquad
\sum_{\ell=\ell_{0}}^{\infty}x^{\ell-\ell_{0}}=\frac{1}{1-x}.
\label{eq:app_geom}
\end{equation}

\paragraph{Even sector ($\varepsilon=+1$).}
Writing $E_{n,\ell,m_{s}}=\omega(1-\vartheta m_{s})+2\omega n+2\omega\ell$:
\[
\sum_{n\geq0}\sum_{\ell\geq\ell_{0}}e^{-\beta E_{n,\ell,m_{s}}}
=\frac{e^{-\beta\omega(1-\vartheta m_{s})}\,x^{\ell_{0}}}{(1-x)^{2}}
=\frac{e^{-\beta\omega(1+2\ell_{0}-\vartheta m_{s})}}{(1-x)^{2}}.
\]
Summing over $m_{s}=\pm1$:
\[
Z(\beta)
=\frac{e^{-\beta\omega(1+2\ell_{0})}
       \bigl(e^{\beta\omega\vartheta}+e^{-\beta\omega\vartheta}\bigr)}
       {(1-x)^{2}}
=\frac{2\,e^{-\beta E_{0}}\cosh(\beta\omega\vartheta)}
       {(1-e^{-2\beta\omega})^{2}},
\]
with $E_{0}=\omega(1+2\ell_{0})$, matching the $\varepsilon=+1$ case of \eqref{eq:E0cases}.

\paragraph{Odd sector ($\varepsilon=-1$).}
The derivation proceeds identically with $\ell\to\ell+\nu$ in the exponent. The geometric sums are unaffected, and one obtains
\[
Z(\beta)
=\frac{2\,e^{-\beta E_{0}}\cosh(\beta\omega\vartheta)}
       {(1-e^{-2\beta\omega})^{2}},\qquad
E_{0}=\omega\bigl(1+2(\ell_{0}+\nu)\bigr),
\]
matching the $\varepsilon=-1$ case of \eqref{eq:E0cases}.

The compact form \eqref{eq:Zgeneral} therefore holds for both reflection sectors, with all dependence on~$\ell_{0}$, $\nu$, and $\varepsilon$ absorbed into the single ground-state energy~$E_{0}$.


\end{document}